\documentclass[12pt,psf,epsf]{article}

\usepackage[centertags]{amsmath}
\usepackage{graphicx}
\usepackage{epsfig}
\usepackage{ulem}
\usepackage[english]{babel}
\usepackage{array}
\usepackage{amsthm}
\usepackage{latexsym}
\usepackage[mathcal]{euscript}
\pdfoutput=1
\usepackage{epsfig}
\usepackage{color}
\usepackage{jheppubm}

\newcommand{\be}{\begin{equation}}
\newcommand{\ee}{\end{equation}}
\newcommand{\bea}{\begin{eqnarray}}
\newcommand{\eea}{\end{eqnarray}}

\newcommand{\bi}{\begin{itemize}}
\newcommand{\ei}{\end{itemize}}

\DeclareMathOperator\const{const}

\title{Orientation Dependence  of
  Confinement-Deconfinement Phase Transition in Anisotropic Media}

\author{Irina Aref'eva$^a$, Kristina Rannu$^{b}$ and  Pavel Slepov$^{c}$}

\affiliation{$^a$Steklov Mathematical Institute, Russian Academy of
  Sciences, Gubkina str.8, 119991, Moscow, Russia\\
  $^b$Peoples Friendship University of Russia, Miklukho-Maklaya
  str.6, 117198, Moscow, Russia\\
  $^c$Moscow State University, Faculty of Physics, 1-2 Leninskie
  Gory, 119991, Moscow, Russia}

\emailAdd{arefeva@mi.ras.ru}
\emailAdd{rannu\underline{~}ka@pfur.ru}
\emailAdd{slepov@mi.ras.ru}

\abstract{We study the $T$-$\mu $ phase diagram of anisotropic media,
  created in heavy-ion collisions (HIC). Such a statement of the
  problem is due to several indications that this media is anisotropic
  just after HIC. To study $T$-$\mu $ phase diagram we use holographic
  methods. To take into account the anisotropy we use an anisotropic
  black brane solutions for a bottom-up QCD approach in 5-dim
  Einstein-dilaton-two-Maxwell model constructed in our previous
  work. We calculate the minimal surfaces of the corresponding probing
  open string world-sheet in anisotropic backgrounds with various
  temperatures and chemical potentials. The dynamical wall (DW)
  locations, providing the quark confinement, depend on the
  orientation of the quark pairs, that gives a crossover transition
  between confinement/deconfinement phases in the dual gauge theory.

}

\begin{document}
\maketitle

\section{Introduction}

Study of the phase diagram in the temperature and chemical potential
$(\mu , T)$-plane is one of the most important questions for QCD
\cite{1710.09425}. It aims to describe the physics of strongly
interacting matter at extreme energy densities, where we have evidence
that a new phase of matter, the quark gluon-plasma (QGP) appears, as
well to understand the matter, which prevailed in early Universe in
first 30 micro seconds.

It is well known that perturbative methods are inapplicable to study
this subject. The lattice QCD still has difficulties with the study of
theories with non-zero chemical potential due to the sign problem
\cite{Aarts}. The gravity/gauge duality provides an alternative tool to
study $T$-$\mu $ phase diagram \cite{Solana,IA,DeWolf}.

The phase diagram has been experimentally studied only for small
$\mu $ and large $T$ values (RHIC, LHC) on the one hand and for low
energies (small $T$) and finite chemical potential values (SPS) on the
other hand. The experimental study of the phase diagram in between these
two particular cases is one of the main tasks of FAIR and NICA, now
being under construction. For this purpose the results of the beam
scanning in HIC are supposed to be analyzed. In this context note, that
there is an obvious anisotropy (the nonequivalence of the longitudinal
and transverse directions) in the substance produced in HIC. In fact it
is believed, that QGP formed in HIC is initially in an anisotropic state
and isotropization occurs approximately in $0.5 \div 2$~fm/c after a
collision \cite{Strickland:2013uga}. Therefore it seems natural
to assume that the results of the beam scanning will study the phase
transition in an anisotropic QCD (with parameter of anisotropy depending
on time). One cannot use the anisotropic lattice QCD
\cite{Karsch82,Karsch89,Arefeva:1993rz,Forcrand2018} to study $T$-$
\mu $ phase diagram because of the well known sign problem mentioned
above.

This anisotropy of QGP can be taken into account holographically. An
additional argument to use anisotropic holographic model is that it
supports the estimation of multiplicity \cite{AGG}. We use the
bottom-up holographic model to study $T$-$\mu $ phase diagram and
investigate the anisotropy influence on it. On the gravity side
anisotropy is supplied with the magnetic ansatz of Maxwell field to
dilaton gravity action. Non-zero chemical potential is introduced via
electric ansatz for the second Maxwell field
\cite{ArefevaRannu}. Thereby the 5-dimensional dilaton gravity with
two Maxwell fields turns out to be the most suitable model. Such model
was considered in \cite{ArefevaRannu,Arefeva:2016rob}. The simplest
anisotropic model, characterized by anisotropic parameter $\nu $, has
been investigated in \cite{AGG}. The feature of this particular
model is that it correctly reproduces the energy dependence of
multiplicity of charged particles produced at LHC in heavy ion
collisions (results by ATLAS and Alice \cite{ATLAS,ALICE}). Other
holographic models cannot fit experimental data in spite of several
attempts (see our previous papers on the subject
\cite{AGG,IA-Q2018,ArefevaRannu}). We take $\nu = 4.5$ to fit the
experimental data.

The specific view of $T$-$\mu $ diagram describing the
confinement/deconfinement phase transition in anisotropic media depends
on orientation of the quark pair relative to the anisotropy axis.
Anisotropy axis in QGP created in HIC is defined by the axes of ions
collisions. In our previous work we studied longitudinal and transverse
orientation cases and showed that they have two different $T$-$\mu $
curves on phase diagram \cite{ArefevaRannu}. But it is obvious
that in real experiment, especially for large chemical potentials, the
quark pair orientation should be random thus causing blurring of the
phase transition line. In this paper we investigate the general
confinement/deconfinement phase transition picture for arbitrary quark
pair orientation and describe the emerging variety of scenarios.

We consider a 5-dim metric defined by anisotropic parameter
$\nu $, non-trivial warp-factor, non-zero time component of the first
Maxwell field and non-zero longitudinal magnetic component of the second
Maxwell field. We take the warp-factor in the simplest form
$b(z) = e^{\frac{c z^{2}}{2}}$, as this particular case allows to
construct explicit solution \cite{ArefevaRannu}. We study the
confinement/deconfinement phase transition line for the pair of quarks
in the anisotropic QGP. We show the dependence of the
confinement/deconfinement phase transition on the angle $\theta $ between quarks
line and heavy-ion collisions line. We calculate the expectation values
of the rectangular temporal Wilson loop $W_{\theta T}$ for different
orientation of the spacial part of the Wilson loop and find the
conditions of the confinement/deconfinement phase transition for this
line. For this purpose we introduce the effective potential
$\mathcal{V}(z)$, that depends on the angle $\theta $ and describes the
interquark interaction. The confinement takes place, when the effective
potential $\mathcal{V}$ has a critical point. We find conditions, under
which the critical point exists, and study the dependence of the
confinement/deconfinement phase transition temperature on chemical
potential $\mu $ and angle $\theta $.

The specific feature of the holographic description of the
confinement/deconfinement is the position of the phase diagram
associated with the Wilson loop behavior relative to the line of the
Hawking-Page phase transition, characterized by the 5-dim background
metric. It is evident, that unlike the confinement/deconfinement
transition line, the Hawking-Page transition line's position on the
phase diagram doesn't depend on the angle $\theta $. As a result the
change of this angle leads to changing of the mutual arrangement of the
confinement/deconfinement transition line and Hawking-Page transition
line on the phase diagram. We find the critical value, for
which the top of the Hawking-Page transition line, corresponding to
$\mu = 0$, and the top of the confinement/deconfinement transition line
coincide.

The paper is organized as follows. In Sect.~\ref{setup} we briefly
describe the 5-dim black brane solution in the anisotropic background
(Sect.~\ref{model}) and sketch calculates of the expectation value of the
temporal Wilson loop (Sect.~\ref{wilson-loop}). In Sect.~\ref{conf-deconf}
we find the condition of the confinement-deconfinement phase transition
for zero and non-zero temperature. In Sect.~\ref{results} we perform
detailed phase diagrams depending on the angle $\theta $ and in
Sect.~\ref{discussion} discuss further directions of investigation of
holographic anisotropic QCD.

\section{Setup}%
\label{setup}

\subsection{The model}%
\label{model}
We consider a 5-dimensional Einstein-dilaton-two-Maxwell system. In the
Einstein frame the action of the system is specified as
\begin{gather}
  S = \int \frac{d^{5}x}{16\pi G_{5}} \, \sqrt{-\det (g_{\mu \nu })}
  \left[ R - \frac{f_{1}(\phi )}{4} \ F_{(1)}^{2} -
    \frac{f_{2}(\phi)}{4} \ F_{(2)}^{2} - \frac{1}{2} \ \partial_{\mu}
    \phi \partial^{\mu} \phi - V(\phi) \right], \label{action.eq1}
\end{gather}
where $F_{(1)}^{2}$ and $F_{(2)}^{2}$ are the squares of the Maxwell
fields $F_{\mu \nu }^{(1)} = \partial _{\mu } A_{\nu }^{(1)} -
\partial _{\nu } A_{\mu }^{(1)}$ and $F_{\mu \nu }^{(2)} = q \ dy^{1}
\wedge dy^{2}$, $f_{1}(\phi )$ and $f_{2}(\phi )$ are the gauge kinetic
functions associated with the corresponding Maxwell fields,
$V(\phi )$ is the potential of the scalar field $\phi $.

\begin{figure}[t!]
  \centering
  \includegraphics[scale=0.60]{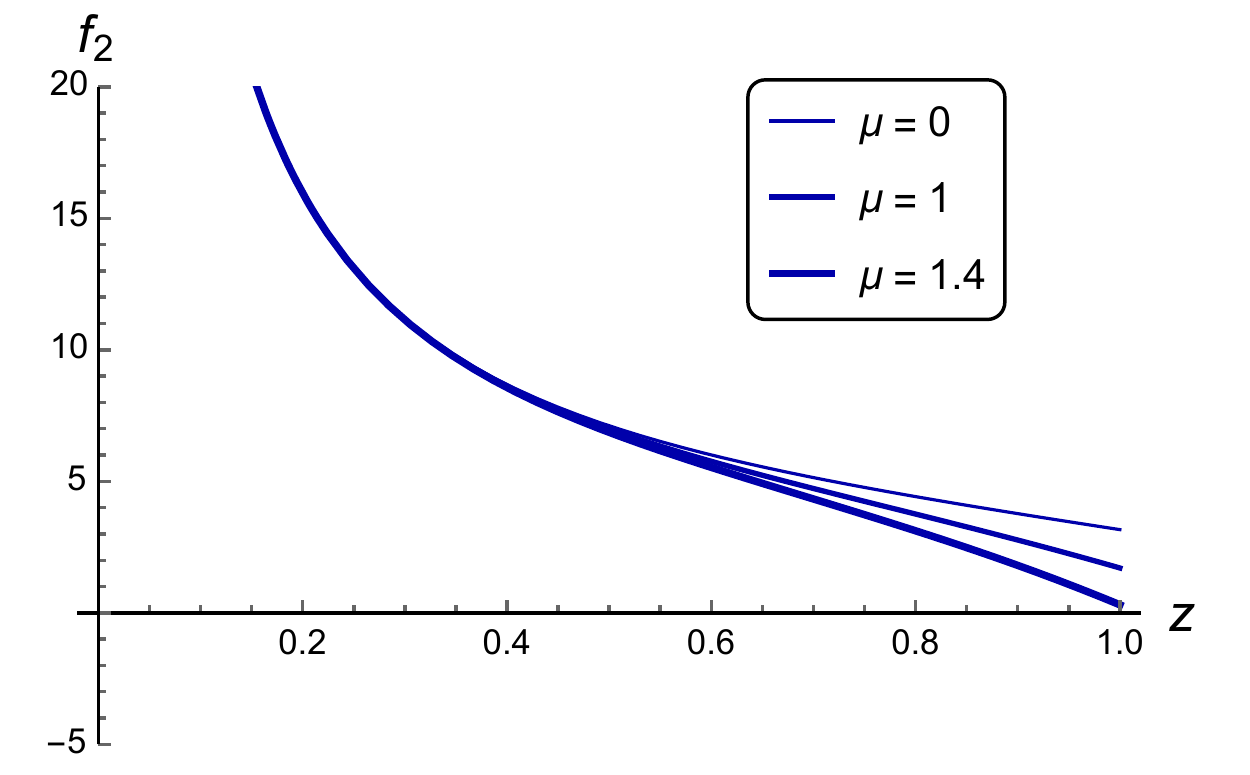}A \hspace{2cm}
  \includegraphics[scale=0.40]{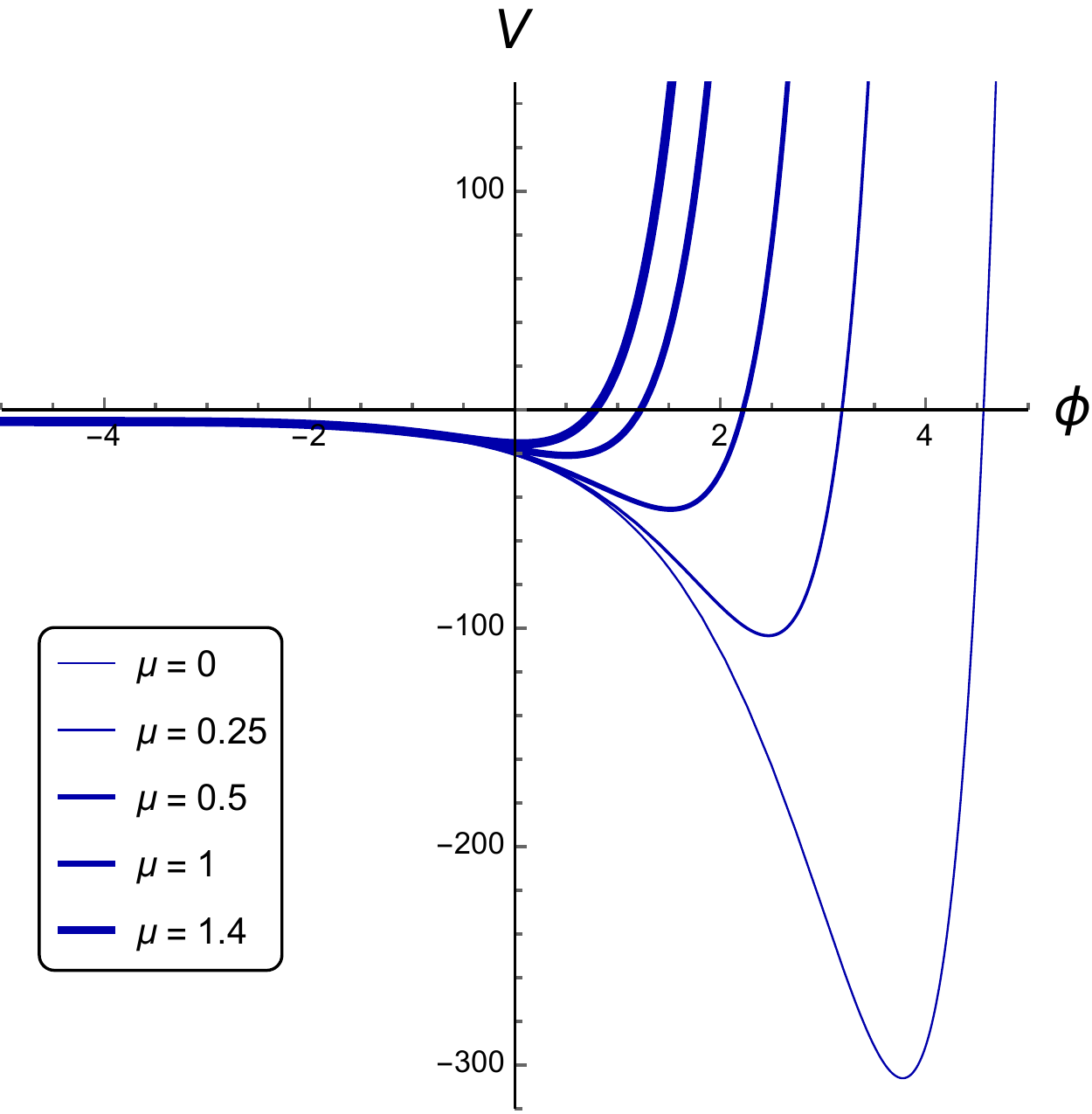}B
  \caption{A) Coupling function $f_2(z)$  for $z_h = 1$, $c = - 1$, $q =
    1$ and different $\mu$ in anisotropic case, $\nu = 4.5$. B) Scalar
    field potential $V(\phi)$  for $z_h = 1$, $c = -1$ and different
    $\mu$ in anisotropic case, $\nu = 4.5$.}
  \label{Fig:V-approx-f}
\end{figure}

To find the black brane solution in the anisotropic background we used
the metric ansatz in the following form:
\begin{gather}
  ds^{2} = G_{\mu \nu}dx^{\mu}dx^{\nu} \frac{L^{2} \, b(z)}{z^{2}}
  \left[ - \ g(z) dt^{2} + dx^{2} + z^{2-\frac{2}{\nu}} \left(
      dy_{1}^{2} + dy_{2}^{2} \right) + \frac{dz^{2}}{g(z)}
  \right], \label{eq:2.02} \\
  \phi = \phi (z), \qquad
  A_{\mu }^{(1)} = A_{t} (z) \delta _{\mu}^{0}, \qquad
  F_{\mu \nu }^{(2)} = q \ dy^{1} \wedge dy^{2}, \label{eq:2.03}
\end{gather}
where $b(z)$ is the warp factor and $g(z)$ is the blackening function
(see (2.31) and (2.38) in \cite{ArefevaRannu}); we set the AdS
radius \mbox{$L = 1$}. The coupling function $f_{2}$, directly connected with
the model anisotropy, depends on chemical potential as well
(Fig.~\ref{Fig:V-approx-f}.A). The potential $V$
(Fig.~\ref{Fig:V-approx-f}.B) can be approximated by a sum of two
exponents and a negative constant (see (2.70)--(2.74) in \cite{ArefevaRannu}). Functions $f_{1}$ and $f_{2}$ are given by eqs. (2.17) and
(2.51) in \cite{ArefevaRannu}. All the quantities in formulas and
figures are presented in dimensionless units.

\subsection{The Wilson loop}%
\label{wilson-loop}

The purpose of our consideration is to calculate the expectation value
of the temporal Wilson loop
\begin{gather}
  W[C_{\theta}] = e^{-S_{\theta,T}},
\end{gather}
oriented along vector $\vec{n}$, such that $n_{x} = \cos \theta $,
$n_{y} = \sin \theta $.

\begin{figure}[b!]
  \centering
  \includegraphics[scale=0.25]{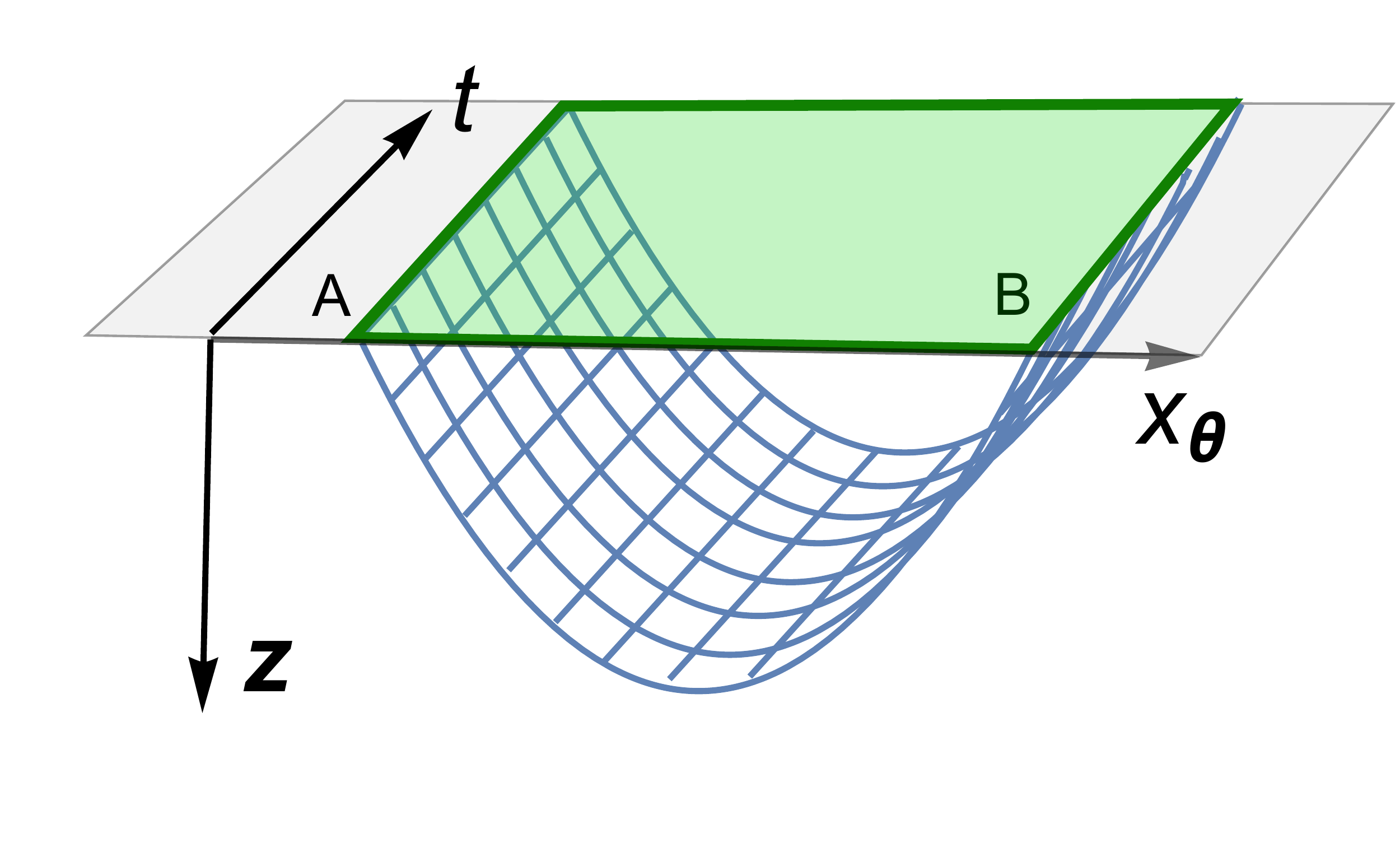}A
  \includegraphics[scale=0.25]{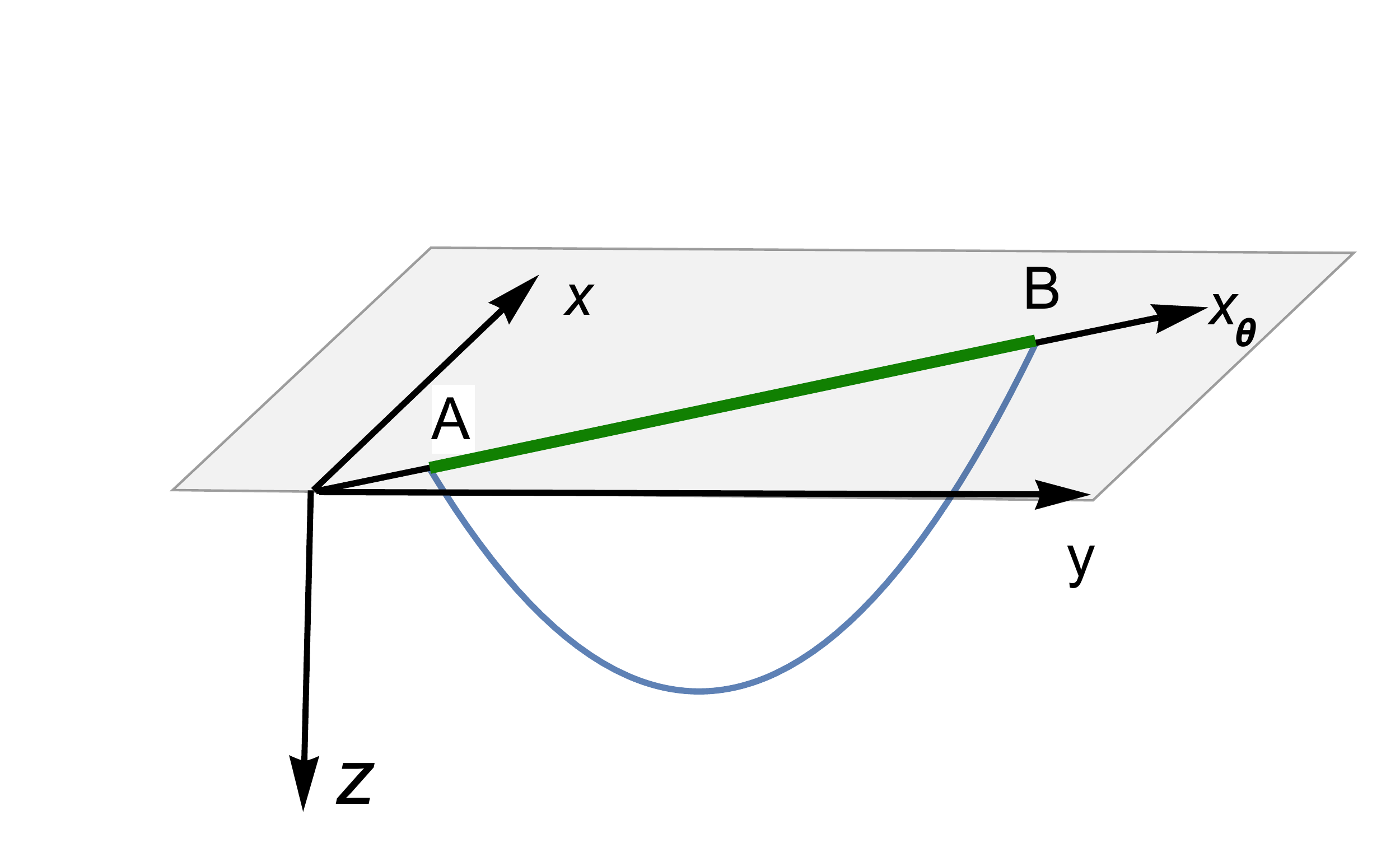}B
  \caption{A) Wilson loop and the world sheet. B) Projection of the
    world sheet to  fixed $t$.}
  \label{Fig:WL}
\end{figure}
Following the holographic approach \cite{Maldacena, Rey,
  J.Sonnenschein} we have to calculate the value of the Nambu-Goto
action for the test string in our background:
\begin{gather}
  S = - \ \frac{1}{2 \pi \alpha '} \int d t \, d\xi \, e^{\sqrt{
      \frac{2}{3}}\phi (z)} \sqrt{-\det G_{\mu \nu } \partial _{\alpha
    } X ^{\mu } \partial _{\beta } X^{\nu }}, \label{action}
\end{gather}
where $G_{\mu \nu }$ is given by \eqref{eq:2.02}. The world sheet
presented in Fig.~\ref{Fig:WL} is parameterized as
\begin{gather*}
  X^{0} \equiv t, \quad
  X^{1} \equiv x = \xi \cos \theta , \quad
  X^{2} \equiv y_{1} = \xi \sin \theta , \quad
  X^{3} \equiv y_{2} = \const, \quad
  X^{4} \equiv z = z(\xi).
\end{gather*}
The action \eqref{action} can be rewritten:
\begin{gather}
  S = - \ \frac{\tau }{2 \pi \alpha'} \int d\xi \ M(z(\xi))
  \sqrt{\mathcal{F}(z(\xi)) + (z'(\xi))^{2}}, \quad
  \tau = \int d t, \label{eq:3.01} \\
  M(z(\xi)) = \frac{ b(z(\xi))}{z(\xi)^{2}}\ e^{\sqrt{\frac{2}{3}}\phi
    (z)}, \quad
  \mathcal{F}(z(\xi)) = g(z(\xi)) \left (z(\xi )^{2-\frac{2}{\nu}}
    \sin^{2}(\theta) + \cos^{2}(\theta)\right).
\end{gather}
Let us introduce the effective potential:
\begin{gather}
  {\mathcal{V}}(z) \equiv M(z) \sqrt{\mathcal{F}(z)}. \label{Poten}
\end{gather}
From {\eqref{eq:3.01} we have representations for the character length
of the string and the action:
\begin{gather}
  \frac{\ell}{2} = \int_{0}^{z_{*}} \frac{1}{\sqrt{\mathcal{F}(z)}} \
  \frac{dz}{\sqrt{\cfrac{\mathcal{V}
        ^{2}(z)}{\mathcal{V}^{2}(z_{*})}-1}}, \label{ell} \\
  \frac{{S}}{2} = \int_{\epsilon }^{z_{*}}
  \frac{\mathcal{V}(z)}{\mathcal{V}(z_{*})} \
  \frac{M(z)dz}{\sqrt{\cfrac{
        \mathcal{V}^{2}(z)}{\mathcal{V}^{2}(z_{*})}-1}}, \label{calS}
\end{gather}
where $z_{*}$ is a top point. Here we introduce the UV cut-off
$\epsilon $ since $M$ has singular behavior near $z\sim 0$:
\begin{gather}
  M(z) \underset{z\sim 0}{\sim } \frac{M_{0}}{z^{k}}, \qquad k \ge 1.
\end{gather}
From \eqref{ell} and \eqref{calS} we see that $S$ and $\ell $ make sense
if the potential is a decreasing function in the interval $0 < z < z_{*}$,
\begin{gather}
  {\mathcal{V}}(z) > \mathcal{V}(z_{*}), \quad 0 < z_{*} < z_{min},
\end{gather}
where $z_{min}$ is the local minimum of $\mathcal{V}(z)$, $z_{min}<z
_{h}$. We are interested in studying the asymptotics of $S$ at large
$\ell $. To get $\ell \to \infty $ and $S \to \infty $ we have to take
$z_{*} = z_{min}$. Indeed, substituting
\begin{gather}
  \frac{ \mathcal{V}^{2}(z) }{ \mathcal{V}^{2}(z_{min}) } = 1 +
  \mathcal{V}_{2} (z - z_{min})^{2} + o((z-z_{min})^{2}), \qquad
  \mathcal{V}_{2} \equiv \frac{\mathcal{V}''(z_{min})}{\mathcal{V}(z _{min})},
\end{gather}
into \eqref{ell} and \eqref{calS}, we get
\begin{gather}
  \ell = 2 \int _{0}^{z_{min}} \frac{dz}{\sqrt{ \mathcal{F}(z) \,
      \mathcal{V}_{2} } \ (z_{min}-z)}  \sim
  \sqrt{\frac{{\mathcal{V}(z _{min})}}{\mathcal{F}(z_{min})\
      \mathcal{V}''(z_{min})}} \ \log (z_{min}-z), \\
  S = 2 \int _{\epsilon }^{z_{min}}
  \frac{\mathcal{V}(z)M(z)dz}{\mathcal{V}(z_{min})\sqrt{ \,
      \mathcal{V}_{2} } \ (z_{min}-z)}  \sim M(z_{min})\sqrt{
    \frac{\mathcal{V}(z_{min})}{\mathcal{V}''(z_{min})}}\ \log
  (z_{min}-z),
\end{gather}
so that $\ell \rightarrow \infty $ as $z \rightarrow z_{min}-0$ and
$S \rightarrow \infty $ as $z \rightarrow z_{min} - 0$.

The stationary point, $\mathcal{V}^{\prime }|_{z=z_{min}} = 0$, is
usually called a dynamical wall (DW) point and satisfies the equation:
\begin{gather}
  z = z_{DW}\!: \ \frac{M'(z)}{M(z)} + \frac{1}{2}
  \frac{\mathcal{F}'(z)}{ \mathcal{F}(z)} = 0. \label{EQPoten}
\end{gather}
Taking the top point $z_{*} = z_{DW}$, we get
\begin{gather}
  S \sim \sigma _{DW}\ \ell , \qquad
  \sigma _{DW} = M(z_{DW}) \,\sqrt{\mathcal{F}(z_{DW})}.
\end{gather}

\section{Confinement/deconfinement phase transition}%
\label{conf-deconf}

\begin{figure}[h!]
  \centering
  \includegraphics[scale=0.38]{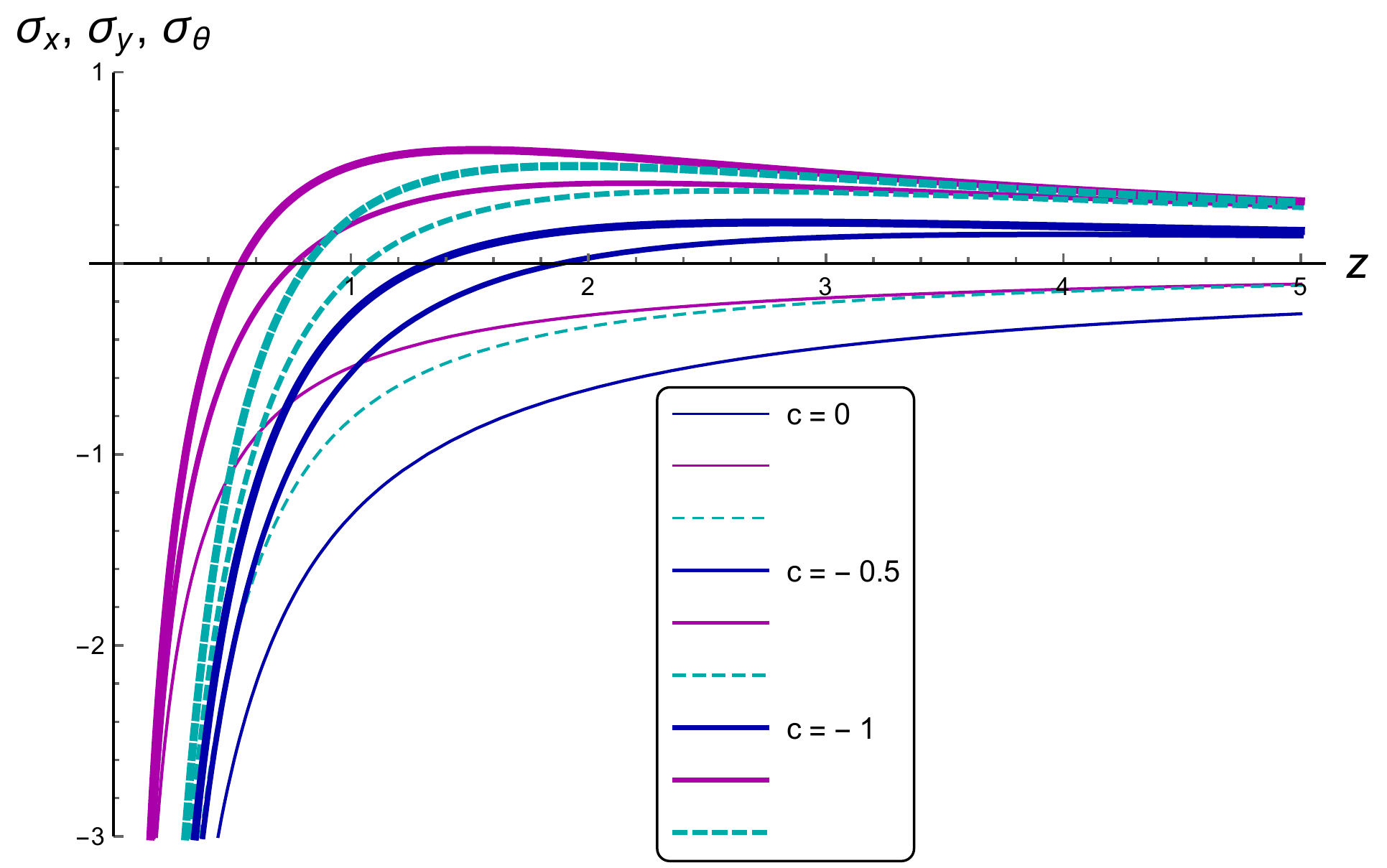}A
  \includegraphics[scale=0.38]{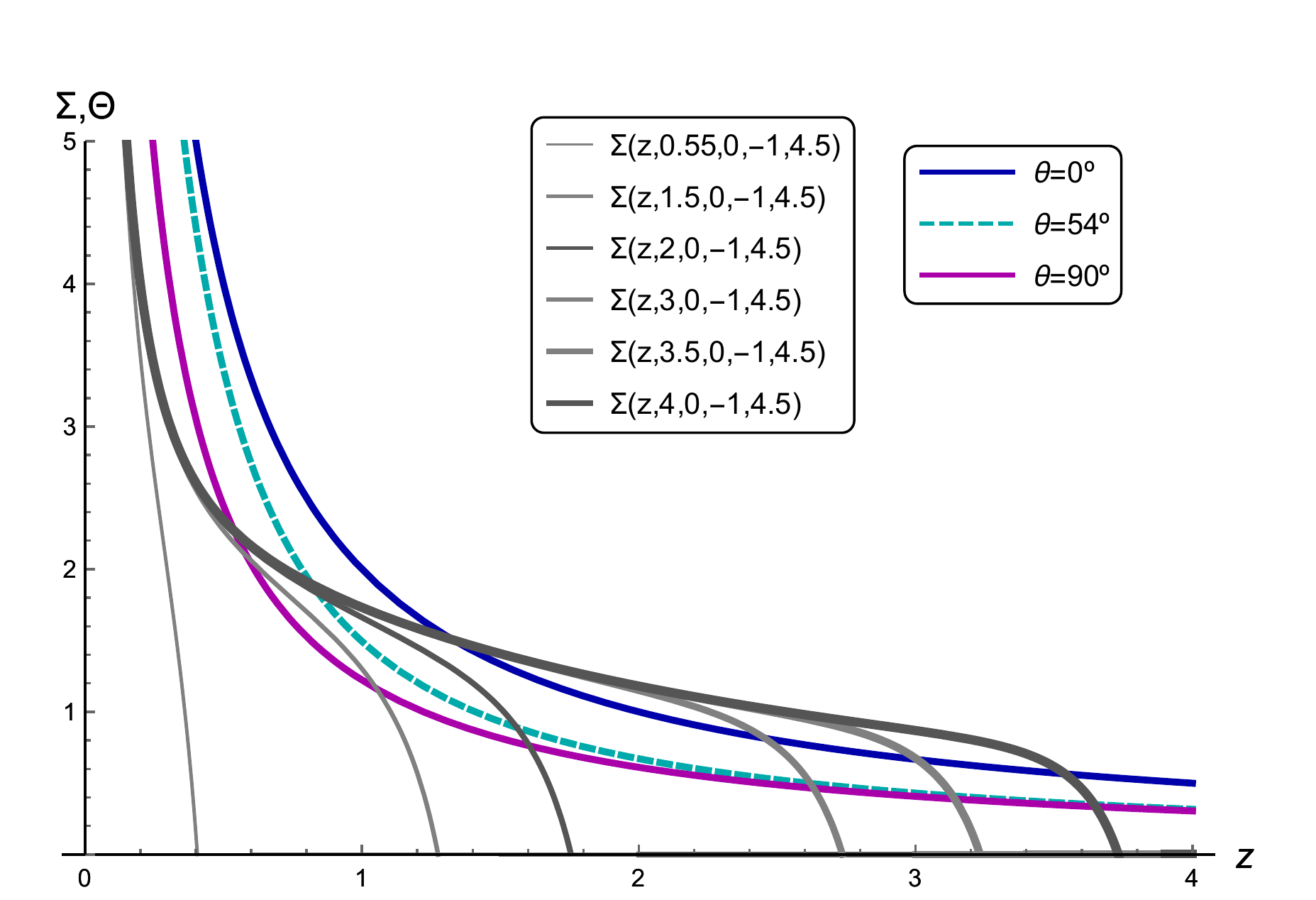}B
  \caption{A) Functions $\sigma_x (z,c,\nu)$ (blue lines),
    $\sigma_y (z,c,\nu)$ (magenta lines),
    $\sigma_{\theta} (z,c,\nu,\theta)$ (cyan dashed lines) for the
    angle $\theta = 54^{\circ}$, $\nu = 4.5$ and for different $c$. B)
    DWs' positions, corresponding to the Wilson lines $W_{xT}$,
    $W_{yT}$, $W_{\theta T}$ in the anisotropic case $\nu = 4.5$, are
    given by intersections of gray lines representing $\Sigma(z)$ and
    the blue, cyan, magenta lines representing another part of the
    equation \eqref{sig} for angles $\theta = 0^{\circ}, 54^{\circ},
    90^{\circ}$. Here we vary $z_h$ and $c$. In all cases to get the
    DW position we take the minimal intersection point.}
  \label{fig:image4}
\end{figure}
In our case the effective potential depends on the warp factor, the
scalar field and the angle. To find stationary points of
$\mathcal{V}(z)$ we solve the equation \eqref{EQPoten} for the
potential \eqref{Poten} with arbitrary angle. This equation has the form
\begin{gather}
  \Sigma (z,z_{h},\mu ,c,\nu ) - \Theta (z,\nu ,\theta ) = 0, \label{sig}
\end{gather}
where $\Sigma (z,z_{h},\mu ,c,\nu )$ does not depend on $\theta $:
\begin{gather}
  \Sigma (z,z_{h},\mu ,c,\nu ) \equiv \sigma (z,c,\nu )+\frac{g'(z)}{2
    g(z)}, \\
  \sigma (z,c,\nu )\equiv c z + \frac{1}{\nu z} \, \sqrt{\frac{2}{3}} \,
  \sqrt{3 c \, \nu ^{2} z^{2} \left ( \frac{c z^{2}}{2} - 3 \right ) + 4 \nu -
    4}, \nonumber
\end{gather}
and $\Theta (z,\nu ,\theta)$ does:

\begin{gather}
  \Theta (z,\nu,\theta) \equiv \frac{2}{z} - \frac{\left( 1 -
      \frac{1}{\nu} \right) z^{1-\frac{2}{\nu}} \sin
    ^{2}(\theta)}{\cos^{2}(\theta) + z^{2-\frac{2}{\nu}} \sin^{2}(\theta)}.
\end{gather}
Equations for DWs in particular cases for $\theta = 0, \, \pi /2$,
considered previously in \cite{ArefevaRannu}, follow immediately
from equation \eqref{sig}. For zero temperature, i.e. $g = 1$, we get
equation
\begin{gather}
  \sigma_{\theta }(z,c,\nu ) \equiv \sigma (z,c,\nu ) -\Theta (z,\nu,
  \theta ) = 0. \label{LHStheta}
\end{gather}

The existence/non-existence of solutions of equations \eqref{LHStheta}
and \eqref{sig} can be seen graphically at Fig.~\ref{fig:image4}. At
Fig.~\ref{fig:image4}.A plots of $\sigma_{x} = \sigma _{\theta =0}$,
$\sigma _{y} = \sigma _{\theta =\pi /2}$ and $\sigma _{\theta }
(z,c,\nu ,\theta )$ for a particular $\theta = 54^{\circ}$ are
presented. We see that for $c = 0$ there are no roots of equation
\eqref{LHStheta}, meanwhile there are roots for any orientations of
the Wilson line and negative $c$ presented here.

For non-zero temperature it is more convenient to find the roots of
equation \eqref{sig} by drawing $\Sigma (z,z_{h},\mu ,c,\nu )$ and
$\Theta (z,\nu ,\theta )$ separately and find their intersections. In
this case the location of $\Theta (z,\nu ,\theta )$ depends on the
geometry of quark-antiquark pair, meanwhile the location of $\Sigma
(z,z_{h},\mu ,c,\nu )$ depends on $z_{h}$, $\mu $, $c$,~$\nu $. We see
that for these parameters fixed the intersection of $\Sigma
(z,z_{h},\mu ,c,\nu )$ and $\Theta (z,\nu ,\theta )$ depends on the
orientation. In particular, the line $\Sigma (z,1.5,0,-1,4.5)$ at
Fig.~\ref{fig:image4}.B intersects the magenta line and does not
intersect the dashed darker cyan and blue lines. This shows that for
$\theta = 90^{\circ}$ confinement occurs, but for $\theta = 54^{\circ
}$ and, moreover, for $\theta = 0^{\circ }$, does not. The
disappearance of the root we interpret as disappearance of
confinement. The parameters, at which this occurs, define the location
of the Wilson confinement/deconfinement transition line. Since we are
interested in location of the confinement/deconfinement line in the
$(\mu , T)$-plane, we use the expression for the temperature
$T(z_{h},\mu ,c,\nu )$ given by formula (3.1) in
\cite{ArefevaRannu}. The locations of the Wilson
confinement/deconfinement transition lines on the $(\mu , T)$-plane
are presented at Fig.~\ref{Fig:PT} by solid lines. We see that varying
$\theta $ for chemical potential $\mu $ large enough we get essential
spread of position of the confinement/deconfinement transition lines.

For small chemical potentials the situation is more complicated due to
the Hawking-Page instability of the background \eqref{eq:2.02}
\cite{ArefevaRannu}. If the system cools down with the non-zero
chemical potential less than some critical value $\mu _{cr}$, the
background at the temperature $T_{BB}(\mu)$ undergoes the phase
transition from a large to a small black hole. This is a generalization
of the corresponding effect in the isotropic case
\cite{9902170,1012.1864,1108.2029,1301.0385,yang2015}. For zero $\mu $
the Hawking-Page phase transition takes place at $T_{HP}$, where the
free energy equals zero and the black hole dissolves to
thermodynamically stable thermal gas. The particular value of $T_{HP}$
depends on parameters $c$ and $\nu $. For the isotropic background the
Hawking-Page transition temperature $T_{HP}$ is higher than for the
anisotropic one with the same $c < 0$, also the temperature of the
large/small black hole phase transition in the isotropic case is
higher than in the anisotropic one, i.e. $T_{BB}^{(anis)}(\mu) <
T_{BB}^{(iso)}(\mu)$. The value of the critical chemical potential, up
to which this phase transition exists, in the anisotropic case is
larger compared to the isotropic one, $\mu ^{(anis)}_{cr} > \mu
_{cr}^{(iso)}$. Also in \cite{ArefevaRannu} we have found that the
point $(\mu^{(anis)}_{cr}, T^{(anis)}_{cr})$ for $\nu \to 1$ goes
smoothly to $(\mu_{cr}^{(iso)}, T_{cr}^{(iso)})$. The location of the
Hawking-Page transition line for anisotropic case $\nu = 4.5$ is
presented at Fig.~\ref{Fig:PT} by the dashed red line. This line
starts at $(0,T_{HP})$ and end up at $(\mu
_{cr}^{(anis)},T^{(anis)}_{cr})$.

\section{Results}%
\label{results}

The phase diagram in $(\mu , T)$-plane is in fact defined by the
relative disposition of the Hawking-Page transition line and the
Wilson transition line. In the model we have determined the critical
angles ${\theta _{cr1} = 45^{\circ }}$, ${\theta _{cr2} = 54^{\circ}}$
and ${\theta _{cr3} = 65^{\circ }}$. For the critical angle
$\theta_{cr1} = 45^{\circ }$ the Hawking-Page phase transition line
(red dashed line in Fig.~\ref{Fig:PT}) and the phase transition line,
determined by the Wilson loop (orange line in Fig.~\ref{Fig:PT}), have
only one common point $(\mu_{cr}^{(anis)}, T^{(anis)}_{cr})$ at the
end of the Hawking-Page transition line. In this case for $\mu \leq
\mu_{cr}^{(anis)}$ the whole Hawking-Page line determines the
confinement/deconfinement phase transition. For the angle
${\theta_{cr2} = 54^{\circ }}$ the top point (corresponding to $\mu =
0$) of the Hawking-Page phase transition coincides with the top point
of the Wilson phase transition (gray line in Fig.~\ref{Fig:PT}),
$T_{HP} = T_{\text{\footnotesize top},\theta_{cr2}}$. For $\theta
_{cr3} = 65^{\circ}$ the Hawking-Page phase transition line and phase
transition line, determined by the Wilson loop (red solid line), have
only one common point $(\mu_{cr}^{(anis)}, T^{(anis)}_{cr})$ at the
end of the Hawking-Page phase transition line again. In this case the
whole confinement/deconfinement phase transition line is determined by
the Wilson loop only since this line is located below the Hawking-Page
phase transition line.
\begin{figure}[t!]
  \centering
  \includegraphics[scale=0.95]{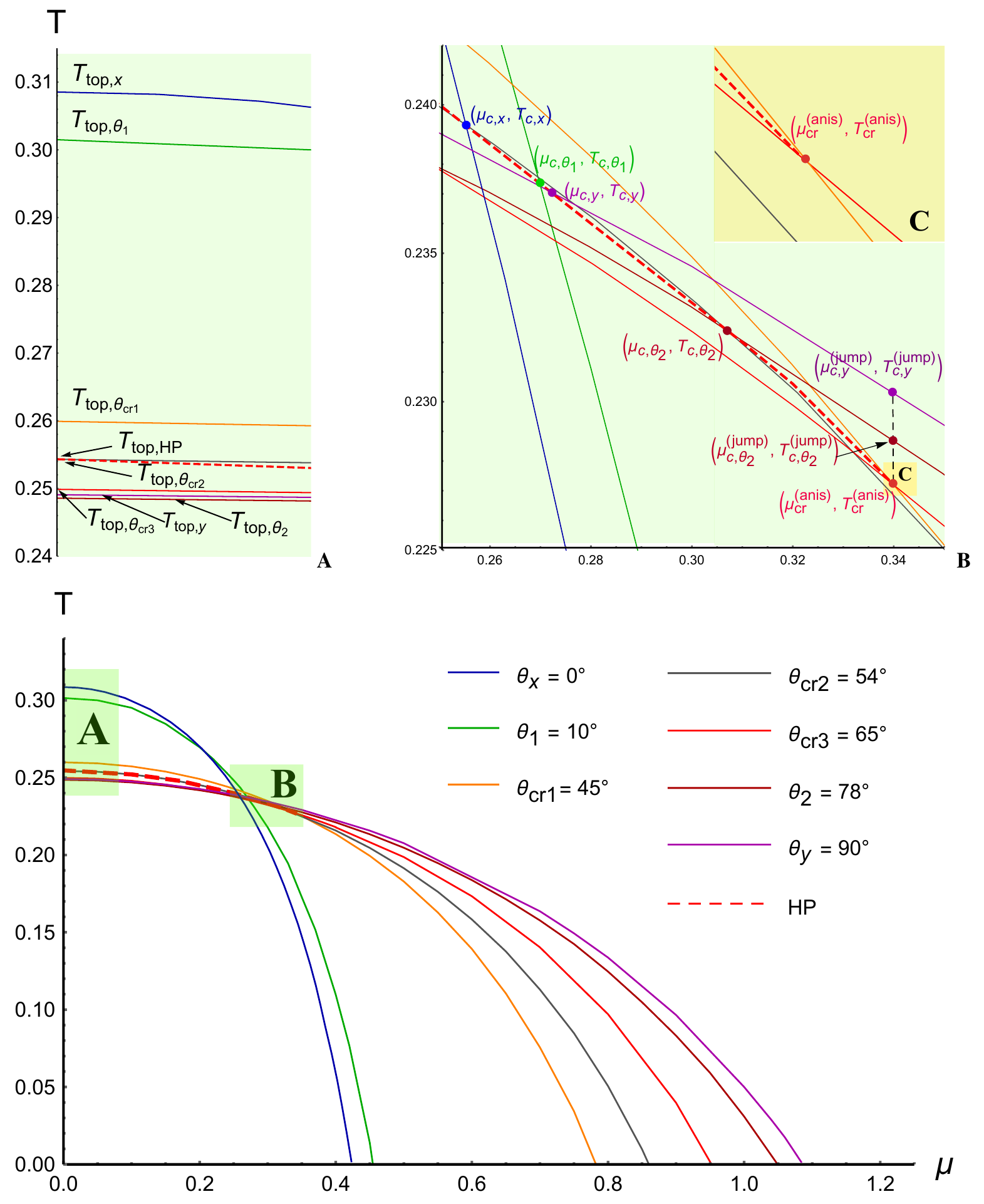}
  \caption{Phase transition diagrams. The plots A) and B) are the zooms of the
    green areas and C) is a zoom of the yellow area in B).}
  \label{Fig:PT}
\end{figure}

Between these critical values of $\theta $ we have the following
pictures.
\begin{itemize}
\item
  For $0^{\circ} \le \theta < 45^{\circ}$ parts of the Wilson
  transition lines near zero values of the chemical potential enter
  the instability regions of our background, where the small black
  holes collapse to large ones. Here the phase transition is
  determined by the Hawking-Page phase transition. After the chemical
  potential exceeds some critical value, the confinement/deconfinement
  phase transition is no longer determined by the background and the
  influence on the Wilson loop starts to dominate, analogous to the
  longitudinal orientation case, presented as $W_{xT}$ in
  \cite{ArefevaRannu} and associated with $\theta_{x} = 0^{\circ}$. The
  green line on Fig.~\ref{Fig:PT} corresponds to $\theta_{1} =
  10^{\circ}$ and shows a typical disposition of the Wilson transition
  line in respect to the Hawking-Page transition line. It intersects
  the Hawking-Page line at the green point $(\mu_{c,\theta_{1}},
  T_{c,\theta_{1}})$ shown at Fig.~\ref{Fig:PT}.B. The intersection of
  the Hawking-Page line with the Wilson transition line corresponding
  to $\theta_x = 0^{\circ}$ is shown by the blue point at
  Fig.~\ref{Fig:PT}.B. While increasing the angle up to $\theta_{cr1}
  = 45^{\circ }$, the intersection reaches the end point of the
  Hawking-Page transition line, $(\mu_{cr}^{(anis)}, T^{(anis)}_{cr})$.
\item
  For $\theta_{cr1} = 45^{\circ}$ the line corresponding to the
  Wilson loop contribution meets the Hawking-Page phase transition
  line at its very end (red point $(\mu_{cr}^{(anis)},
  T^{(anis)}_{cr})$ on Fig.~\ref{Fig:PT}.B).
\item
  For $45^{\circ} < \theta < 54^{\circ}$ we have no intersections of
  the Hawking-Page phase transition line and the Wilson transition
  line lies above, so there is a jump from the Hawking-Page to the
  Wilson transition line.
\item
  For $\theta_{cr2} = 54^{\circ }$ the Wilson line on the phase
  diagram (gray line) starts at $\mu = 0$ at the same point as the
  Hawking-Page transition line, then goes above it and intersects it
  again. Note that the gray line almost coincides with the
  Hawking-Page phase transition line (the dashed red line) at
  Fig.~\ref{Fig:PT}.
\item
  For $54^{\circ} < \theta < 65^{\circ}$ we have no intersections of
  the Hawking-Page phase transition line and the Wilson transition
  line. In this case the whole confinement/deconfinement phase
  transition is determined by the Wilson transition line.
\item
  For $\theta_{cr3} = 65^{\circ}$ the line corresponding to the Wilson
  transition line meets the Hawking-Page phase transition line at its
  very end again (red point $(\mu_{cr}^{(anis)}, T^{(anis)}_{cr})$ on
  Fig.~\ref{Fig:PT}.B).
\item
  For $65^{\circ } < \theta \le 90^{\circ }$ the picture is analogous
  to the case of the transversal orientation case, presented as
  $W_{yT}$ in \cite{ArefevaRannu} and corresponding to $\theta_{y} =
  90^{\circ}$. The confinement/deconfinement phase transition is
  determined by the Wilson transition line starting from the zero
  values of chemical potential up to $\mu_{c,y}$ (magenta point
  $(\mu_{c,y}, T_{c,y})$ on Fig.~\ref{Fig:PT}.B), where it meets the
  instability of the background. Starting from this point the
  Hawking-Page phase transition takes place up to $(\mu_{cr}^{(anis)},
  T^{(anis)}_{cr})$, where we have a jump to a point $(\mu
  _{c,y}^{(jump)}, T_{c,y}^{(jump)})$ on the Wilson transition line. A
  typical Wilson transition line for $\theta_2 = 78^\circ$ is
  presented in darker red. It intersects the Hawking-Page line in
  point $(\mu_{c,\theta_2}, T_{c,\theta_2})$, and the jump from the
  end of the Hawking-Page line $(\mu_{cr}^{(anis)}, T^{(anis)}_{cr})$
  into the darker red point $(\mu_{c,\theta_2}^{(jump)},
  T_{c,\theta_2}^{(jump)})$ is also shown on Fig.~\ref{Fig:PT}.B.
\end{itemize}

\section{Conclusion and discussion}%
\label{discussion}

We have found the dependence of the confinement/deconfinement phase
transition line on the orientation of the quark pair. For this purpose
we have studied the behavior of the temporal Wilson loops in the
particular 5-dimensional anisotropic background supported by dilaton and
two-Maxwell field constructed in \cite{ArefevaRannu}. We
specified the quark pair orientation by angle $\theta$. For each angle
there is its own confinement/deconfinement transition line (see a
variety of these lines in Fig.~\ref{Fig:PT}). There is a Hawking-Page
instability in our background. Combining the phase transition for the
Wilson line with this Hawking-Page instability we have arrived to the
picture presented in Fig.~\ref{Fig:PT}.

At the end we would like to point out, that the possibility of an
experimental check of our estimation of the confinement/deconfinement
line blurring essentially depends on the ability of the experimental
measurement particle yield immediately after HIC. The reason for this
is that the anisotropy effects are expected in the contents of the
fireball, created in HIC, just after collision at times of about
$0.5 \div 2$ fm/c. Anisotropy makes spectrum of the hadrons created by this
fireball depend on the orientation, but this anisotropy disappears very
soon and as a consequence the blurring disappears as well.

As to the future investigations, the following natural questions to
static and non-static properties of our model are worth noting. As has
been mentioned, the anisotropic background constructed in
\cite{ArefevaRannu} can be generalized to provide a more realistic
model. In this case the solution can be given in terms of quadratures
only and we suppose to generalize the Wilson loop calculations to this
more realistic case. As to static properties, it is natural to
\begin{itemize}
\item
  investigate $\theta $-oriented Wilson loops based on more complicated
  factor $b(z)$, in particular such that in the isotropic limit it fits
  the Cornell potential known by lattice QCD;
\item
  study the Regge spectrum for mesons, adding the probe gauge fields to
  the backgrounds and find its dependence on $\theta $;
\item
  consider estimations for direct photons and find dependence on
  orientation \cite{Arefeva:2016rob};
\item
  evaluate transport coefficients and their dependence on the anisotropy;
\item
  estimate the holographic entanglement entropy and find its
  dependence on $\theta $; note that this has been done in \cite{AGG}
  for zero chemical potential and $\theta = 0, \, \pi /2$; the
  isotropic case for non-zero chemical potential has been considered
  recently in \cite{Dudal:2018ztm};
\item
  find a generalization results of \cite{AGP} where an explicit
  isotropic solution for the dilaton potential as a sum of two
  exponents and zero chemical potential has been found.
\end{itemize}

As to the thermalization processes, which are the main motivations of
our consideration of the anisotropic background (see details in
\cite{Arefeva:2016rob,Arefeva:2016lcz}), it would be interesting to
investigate the behavior of the temporal Wilson loop during
thermalization. This problem for zero chemical potential has been
studied in \cite{Hajilou:2017sxf}. It is also interesting to
generalize the result of paper \cite{AAGG} and consider
thermalization of the spacial Wilson loops for non-zero chemical
potential. This will give the dependence of the drag-forces on the
chemical potential.

\section*{Acknowledgments}

This work was partially (I.A. and P.S.) supported by the ``BASIS''
Science Foundation} (grant No.~18-1-1-80-4).

\end{document}